\documentclass[conference]{IEEEtran}

\ifCLASSINFOpdf

\else

\fi
\usepackage{times}
\usepackage{graphicx}
\usepackage{algorithm}
\usepackage{algpseudocode}
\usepackage{setspace}
\usepackage{mathtools}
\usepackage{amsmath}

\begin{document}

\title{ Watchword-Oriented and Time-Stamped Algorithms for Tamper-Proof Cloud Provenance Cognition}

\author{\IEEEauthorblockN{Asif Imran, Nadia Nahar, Kazi Sakib}
\IEEEauthorblockA{Institute of Information Technology\\University of Dhaka\\
Email: asif.imran.anik@gmail.com, nadia.nahar.iit@gmail.com, sakib@univdhaka.edu}
}

\maketitle

\begin{abstract}
Provenance is derivative journal information about the origin and activities of system data and processes. For a highly dynamic system like the cloud, provenance can be accurately detected and securely used in cloud digital forensic investigation activities. This paper proposes watchword oriented provenance cognition algorithm for the cloud environment. Additionally time-stamp based buffer verifying algorithm is proposed for securing the access to the detected cloud provenance. Performance analysis of the novel algorithms proposed here yields a desirable detection rate of 89.33\% and miss rate of 8.66\%. The securing algorithm successfully rejects 64\% of malicious requests, yielding a cumulative frequency of 21.43 for MR.
\end{abstract}


\IEEEpeerreviewmaketitle

\section{Introduction}

Provenance is derivative journal data that represents factual information about events executed on particular operations at the application layer. Cloud computing is the dynamic provisioning of resources from a shared resource pool and needs provenance to ensure integrity and accountability of cloud services. Provenance information journals of various real life cloud applications is required during digital investigation. At the same time, the cloud provenance needs to be tamper-proof to increase acceptability of the data to Digital Forensic Experts. Hence detection and securing of provenance in the cloud is of significant importance for Security Intelligence.

Capturing provenance for Software as a Service (SaaS) applications is an important research issue, since those are dynamic in nature and have volatile attributes. Also, a large number of virtual and physical machines are involved in the cloud and application layer provenance needs to be collected for those for Cloud Security Intelligence. In addition effectively securing the cloud virtual machine (vm) instance that stores provenance using time-stamp based access control has not been considered for cloud environments \cite{armbrust2010view}. More precisely, the following research issues need to be addressed.

\begin{enumerate}
\item Novel algorithms using watchword capabilities to detect provenance in the cloud at the application level of SaaS.
\item Algorithms to tamper-proof the detected provenance using time-stamp based access control and analyzing performance of those in terms of authorized access acceptance and rejection.
\end{enumerate}

Existing mechanisms for provenance cognition in the cloud are not suitable as large volume of data needs to be analyzed, thereby increasing the overhead of the system and slowing down the cloud processes. Framework for provenance detection in the cloud through matching the patterns of journal files to determine system level provenance has been proposed in \cite{imran2013provintsec}. The provenance was analyzed for detection of malware threats to the system. However, the integrity of the captured provenance has not been ensured through tamper-proofing mechanisms. The importance of ensuring cloud accountability is a prime factor for widespread acceptance of cloud among the public as identified in \cite{pohly2012hi}. Methods to secure the cloud data has been proposed through layer-based encryption mechanism. However, the importance of securing the cloud provenance meta-data to aid in digital forensic investigation has been analyzed to a limited extent.

Based on the research questions identified above, this paper proposes methods to detect provenance at the application layer and securing those through time-stamp based access control. Provenance at the application layer of SaaS consists of data such as access-id, time of access, pages viewed, and files modified in terms write operations. The proposed algorithm captures the provenance described above using watchword capability of cloud journals that triggers data insertions into a specific cloud vm-instance using the provenance rules. The rules are used to capture provenance at the application level of cloud where the SaaS modules are deployed. Active-threaded methodology captures provenance at real time through effective monitoring of the cloud vm-instances.

Algorithms called $ObProv$ and $Auditor$ have been proposed in this paper to ensure provenance cognition and verification at the application layer of the cloud. Six real-life applications namely Inventory Management (IM), Accounting Management (AM), Human Resource Management (HRM), Office Document Management (ODM), Sales Management (SM), Customer Relationship Management (CRM) softwares running on OpenStack cloud were tested for provenance cognition using $ObProv$. The applications were implemented as SaaS in 88 vm-instances. Next $Auditor$ algorithm was implemented on the storage vm-instance that measured the request body in the buffer array and checked constraints through access time-stamp matching.

The performance of the proposed algorithms have been evaluated in real life cloud environment, running the specific applications in commercial use over vm-instances. Provenance were captured for those at a Success Rate (SR) of 89.33\% and Miss Rate (MR) of 8.66\%. The mean Rejection Rate (RR) of 64\% requests achieved using $Auditor$ algorithm out of 357 requests made to access the provenance cloud vm-instance. The cumulative frequency of 21.43 was achieved after empirical analysis. Additionally, 65 requests were considered legitimate and granted access to the stored provenance whereas 292 illegitimate requests were rejected by the $Auditor$ algorithm.

\section{Related Work}
A process aware approach to worm attacks and contamination have been designed, implemented and evaluated in \cite{idika2013probabilistic}. The authors identified the important issue that provenance un-awareness leads to problems in quick and accurate identification of worm attack point \cite{idika2013probabilistic}. Process coloring can be assigned to uniquely identify processes, it is inherited by the child processes and diffused through process actions.

\begin{figure*}[t!]
\centering
\includegraphics[scale =0.60]{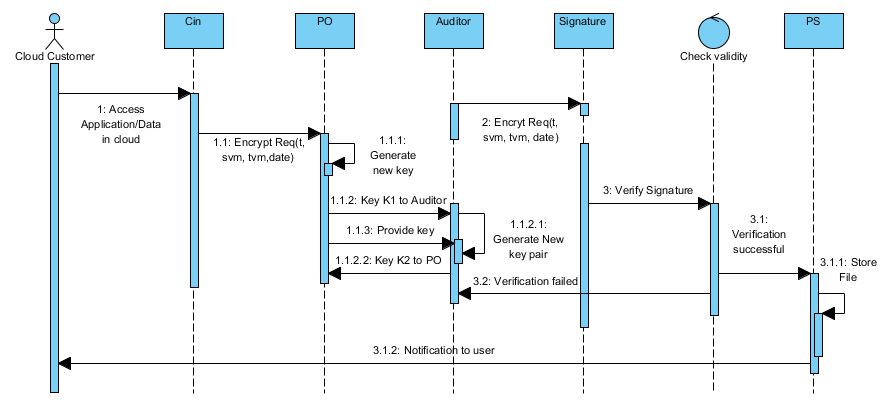}
\caption{Sequence of actions and processes to secure provenance information}
\label{figure3}
\end{figure*}

Process coloring enables fast identification of worm break in point and naturally partitions log data based on colors, thereby ensuring rapid identification \cite{marinho2012provmanager}, \cite{yu2012novel}. The main contribution of the authors is to effectively reduce the volume of log data that need to be analyzed for worm identification. Nevertheless tamper-proofing the captured provenance was unaddressed.

Identified security as an important issue that decelerates the widespread growth of cloud computing \cite{abbadi2011challenges}. Complications for data privacy and data protection have plagued the market of the cloud. The assurance of data integrity is a necessity to ensure acceptance of cloud services in all sectors. The authors suggested a new model that extends the existing model in this regard. However the new model should not threaten the features of the existing model.

Cloud service users need to be vigilant about the security breaches that can occur in the cloud \cite{abbadi2011challenges}. The authors have identified critical security issues due to the nature of service delivery model in the cloud. They have contributed to the cloud research field by identifying critical research questions regarding cloud security. The authors did not identify any malware related vulnerability, neither did their survey bring about any real life scenarios.

The different security and vulnerability assessment in cloud computing environment have been assessed in \cite{ko2011flogger}. Experiments were carried out in three environments that include the applications residing on the same laboratory where the targets are, residing not in the same laboratory but on the same campus, and residing off-campus. Suitable scan tools are applied to the servers to assess what test was applicable. Next the captured vulnerabilities were collected and assessed for risk.

Risk was determined through the relative assessment to the facility in terms of the expected effect on each critical asset \cite{ko2011flogger}. The authors provided experimental evidence on cloud computing on the basis of centralized and de-centralized architectures. The scope of identifying threats through analysis of provenance data have been covered to a limited extent.

\section{Proposed Methododlogy}
Figure \ref{figure3} identifies the flow of sequence to ensure provenance security. The user can access a cloud application through a browser or may request some data file in the cloud storage. The data is stored in the $C_{in}$ and provenance is collected from those using the ProvCapsule algorithm. Next the provenance file containing the time (t), source vm-id (svm), target vm-id (tvm) and date are stored in the provenance file which is then encrypted by PO using its private key. PO generates a private/public key pair and shares it with $Auditor$ process. The proposed $Auditor$ process receives key $K1$ from PO and provides its own public key $K2$ to PO.

The second level of encryption is carried out by the auditor and the provenance file is sent to the signature process for verification. If the time-stamp, source vm, target vm and date matches, it will be forwarded to the Provenance Store, otherwise it will be discarded and the $PO$ and $Auditor$ will both be notified about the issue.

One fundamental difference with the existing research is that the procedures used for signing and verifying provenance information are time consuming since the authorization process is manual \cite{zhang2012track}. However, active-threading technique is proposed here for binding provenance with the original data file in real time, thus the process of forging provenance ownership is prevented in the proposed framework through real-time binding of provenance meta-data to original data. The proposed methodology also reduces time and low-overhead since only critical information like time-stamp, date, target vm and source vm are contained in the provenance file unlike detection of detailed meta-data.

\subsection{ObProv algorithm}

The $ObProv$ algorithm is aimed to encapsulate the provenance information in a specific message body. Hence provenance information produced by $ObProv$ contains only the information needed by the $Auditor$ algorithm or digital forensic experts.

\begin{algorithm}
 \begin{algorithmic}[1]
 \Procedure{ObProv}{$M_{sg},M_{bd}$,$g$}
 \State $M_{pr}\gets Req(T, T_{vm}, S_{vm}, Date)$
 \State $M_{bd}\gets{msg_{1}, msg_{2},.., msg_{n}}$
 \State $Msg\gets{M_{pr}+M_{bd}}$
 \State $A[x]\gets0$
 \State $g\gets 0$
 \While {$g <= sizeof(Msg)$}
 \State $A[g]=Msg$
 \State $g=g+1$
 \State $ObProv\gets ProvOwner{c,d}$
 \EndWhile
 \State If $len.Msg<len.M_{pr}$, then
 \State $Msg\gets len.Req(T, T_{vm}, S_{vm}, Date)$
 \State g = g+1;
 \State \textbf{return} Provenance information file $Msg$
  \EndProcedure
 \end{algorithmic}
 \caption{ObProv algorithm for capturing provenance at the application level}
 \label{euclid}
\end{algorithm}

The $M_{sg}$ is a function that will encapsulate the captured provenance information of ProvOCal into a specific provenance message. At the same time users are allowed to specify what type of provenance information they want to obtain and encapsulate. The $M_{bd}$ consists of the message body. The variable $g$ keeps the length of the size of the provenance information and messages produced by the algorithms, and also acts like a counter. The array $A[x]$ consists of the messages that are stored as strings together with the timestamp $T$. The timestamps are used to compare the authenticity of the requests in the later algorithm. The request body $Req(T,T_{vm},S_{vm},Date)$ ensures that the critical provenance information are specified by the digital forensic experts.

The contents are $T,T_{vm},S_{vm},Date$ to represent time, vm-id, storage allocated for the vm and date of access respectively. Next the message $Msg$ is placed in the array $A[g]$ in the location identified by the current value of $g$. At the same time $ProvOwner$ is read from $ProbCapsule$ and placed in the owner variable of $ObProv$. Finally the $Msg$ is assigned to the $Req$ body and sent to $Auditor$ algorithm for verification.

The algorithms discussed above function in line to detect system level provenance, at the same time those are synchronized to ensure proper activity from provenance cognition, to secured provenance encryption. The next algorithm ensures that unauthorized access to the stored provenance is restricted. Hence there is a sequence of two algorithms that work closely to achieve secured provenance of the big data stored in the cloud.

\subsection{Auditor algorithm}

The $Auditor$ of the process is responsible for verifying the request on the basis of the timestamps. The variable $R_1$ and $R_2$ are used to measure the size of the buffer array and at the same time implement the check constraint that it is under the limit of the initial consideration. Both variables are initialized to 0 to ensure that it can be used as a counter function.

\begin{algorithm}
 \begin{algorithmic}[1]
 \Procedure{Auditor}{$R_1$,$R_2$,$R_3$,$R_4$,$ObProv$,$size$}
 \State $R_1\gets0$
 \State $R_2\gets0$
 \State $Per_i\gets0$
 \State $size\gets1000$
 \State $Buf[size]\gets ObProv$
 \While {$R_1 <= sizeof(Buf[size])$}
 \While {$R_2 <= sizeof(Buf[size])$}
 \State $R_1=sizeof(Buf[size]+Msg)$
 \State $R_2 \gets En.(Req(T, T_{vm}, S_{vm}, Date))$
 \State $R_1=R_1+1$
 \EndWhile
 \EndWhile
 \While {$R_3 <= sizeof(Buffer[size])$}
 \While {$R_4 <= sizeof(Buffer[size])$}
 \State $R_3\gets timetoread.Buffer[size]$
 \State $R_4\gets timeof arrival.Buffer[size]$
 \State If $R_1=R_3$ and
 \State If $R_2!=R_4$
 \State $Per_i=1$
 \State else $Per_i$=0
 \EndWhile
 \EndWhile
  \EndProcedure
 \end{algorithmic}
 \caption{Auditor algorithm for setting permission for provenance access through timestamp authentication}
 \label{euclid}
\end{algorithm}

The two variables $R_3$ and $R_4$ are used to store the timestamps of two consecutive messages. The times of arrival of the new request is matched with the time when the first request if made. At the same time the vm-instance id that is making the request is also recorded. After encryption the provenance information must be stored at the secured end of the trusted Auditor.
\begin{figure*}[t!]
\centering
\includegraphics[scale =0.70]{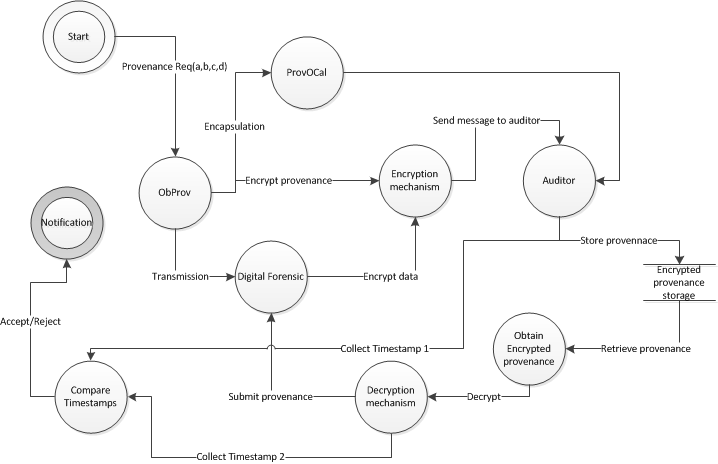}
\caption{Transitions and checks from one state to another in the proposed architecture}
\label{figure4}
\end{figure*}

Algorithms and framework of the trusted model discussed above are explained in this section. At the same time experiments to test the effectiveness of the proposed provenance securing algorithms are conducted with real life cloud environment using OpenStack Essex. Finally $Auditor$ algorithm is implemented at the end of the trusted body to ensure provenance persistency and achieve data integrity.

The two timestamps of the requests do not meet even in the case that the vm-id of those matches since the $Per_i$ will be set to 0. Additionally the permission to access the provenance information will not be allowed because the mismatch of timestamp is caused by a different vm-id philishing the identity of an authorized vm \cite{asif2013cloudniagara}. The difference in the timestamp will ensure that the vm requesting access does not have the authorization and it is a philishing entity.

\subsection{Security goal and assumptions}

The research goal is to prevent the unauthorized or malicious access and nefarious manipulation of provenance information in the cloud. The primary research issues are to determine application level provenance of the identified SaaS in the cloud and ensured secured access for accountability.


The assumptions for the proposed algorithms include that the cloud environment is in Linux platform and the cryptographic algorithms are implemented properly. The forensic investigators who own and maintain the $Auditor$ process is trustworthy. Also, it is assumed that program scripts running at the kernel level are not tampered by the cloud administrator.

\section{Transition of proposed architecture}

The four algorithms discussed above need to communicate with each other using message passing protocols to ensure that data is collected by one algorithm, sent for encryption to the next and finally validated by the remaining algorithms for digital investigation. The proposed algorithmic architecture consists of a number of states that must be transited. At each state proper verification mechanisms need to be implemented to ensure that the process satisfies the requirements needed to advance to the next phase.

\begin{figure*}[t!]
\centering
\includegraphics[scale =0.40]{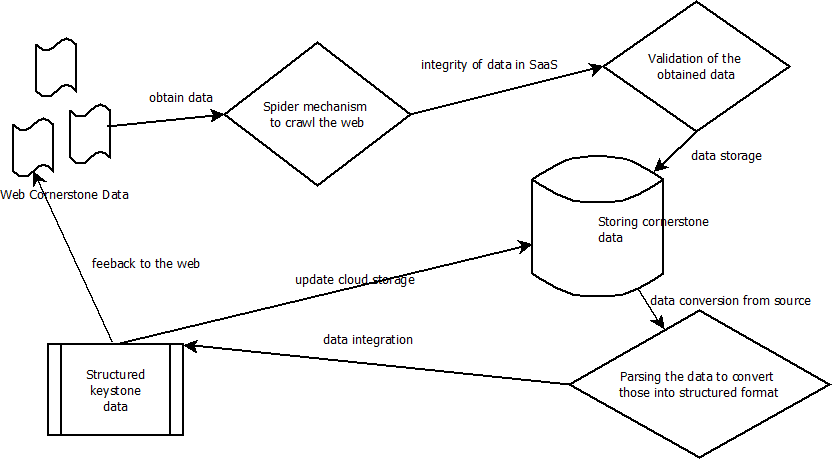}
\caption{Passing captured provenance to the vm-instance with Auditor}
\label{figure8}
\end{figure*}

The start state in Figure \ref{figure4} consists of provenance information being gathered in the request body $Req(a,b,c,d)$ and sent to $ObProv$. Next the request is sent to $ProvCapsule$ so that the collected provenance are encapsulated to its original data file to sid in forensic analysis. Next the bounded provenance and data files are sent to the encryption scheme to obtain the data in encrypted format.

\begin{figure}[t!]
\centering
\includegraphics[scale =0.30]{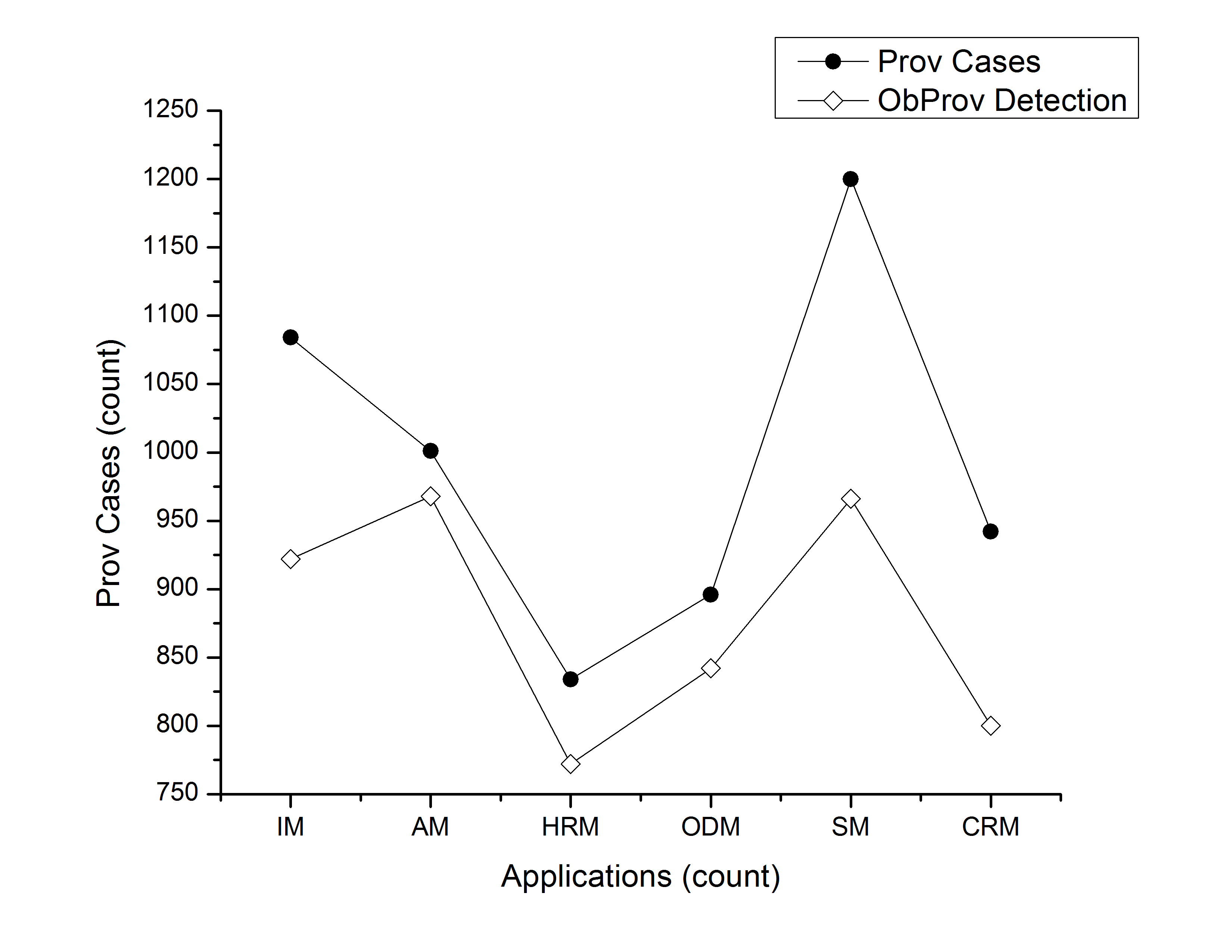}
\caption{Success rate of ObProv compared to real-life provenance cases}
\label{figure9}
\end{figure}

During encryption the provenance owner and $Auditor$ provide public keys and encrypt the information with respective private keys. The hash algorithm is implemented to ensure the integrity of provenance information. In this paper, hashed chain format of encryption is used as opposed to onion type encryption to accommodate the large volume of provenance information as generated by cloud processes and applications.

After encryption, the provenance information is sent to the provenance storage that is maintained by the $Auditor$ since it is assumed that both consumers and service providers consider the $Auditor$ to be a trusted body. During the course of digital investigation if provenance information is required, it is retrieved from the storage using the specific id assigned to provenance data objects by the proposed algorithm. However, the obtained provenance information is still in encrypted form, hence it is decrypted using keys from the $PO$ and $Auditor$ processes for analysis \cite{dastjerdi2012autonomous}, \cite{slipetskyy2011security}.

\section{Analysis of Results}

The performance of the proposed algorithms were tested and analyzed in real life cloud environments running OpenStack Essex cloud. Six commercial real life applications provided as Software as a Service (SaaS) in cloud vm-instances were analyzed at the servers of the cloud service provider. The SaaS applications included Inventory Management Software (IM), Accounting Management Software (AM), Human Resource Management Software (HRM), Office Document Management Software (ODM), Sales Management Software (SM), Customer Relationship Management Software (CRM) as identified in Table I.

\begin{table}
\caption{Performance evaluation of application level provenance cognition }
\begin{tabular}{ l||*{6}{c}r}
\hline
App. & VM-id & Case & Alg Det. & SR (\%) & MR (\%)\\
\hline
IM & 10 & 1084 & 922 & 85.0554 & 14.9446\\
AM & 13 & 1001 & 968 & 96.7032 & 3.2968\\
HRM & 07 & 834 & 772 & 98.3871  & 1.6129\\
ODM & 16 & 896 & 842 & 93.9732 & 6.0268\\
SM & 22 & 1200 & 966 & 80.5000  & 19.5000\\
CRM & 20 & 942 & 800 & 84.9257 & 15.0743\\
\hline
\end{tabular}
\end{table}

\begin{figure}[t!]
\centering
\includegraphics[scale =0.30]{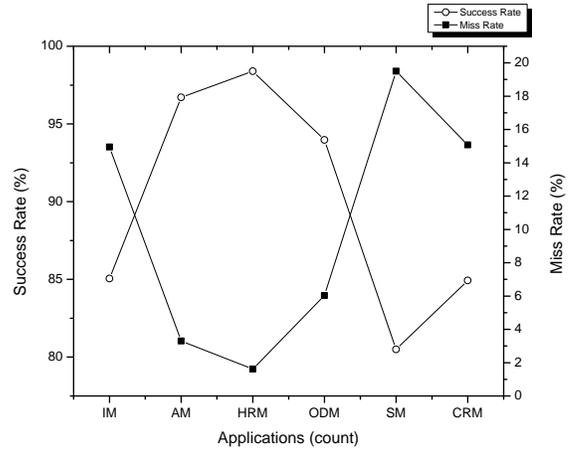}
\caption{Illustration of the high SR and low MR in the proposed architecture}
\label{figure11}
\end{figure}

\begin{table}
\caption{Mean Rejects and Cumulative frequency values for $ObProv$}
\begin{tabular}{ l||*{6}{c}r}
\hline
Req. Size & Count & Acc. & Rej. & M.Delay & C.freq \\
\hline
10-100 & 63 & 13 & 50 & \\
101-200 & 74 & 4 & 70 & \\
201-300 & 69 & 20 & 49 &  64 & 21.43\\
301-400 & 82 & 18 & 64 & \\
401-500 & 69 & 10 & 59 & \\
\hline
\end{tabular}
\end{table}

\begin{figure}[t!]
\centering
\includegraphics[scale =0.30]{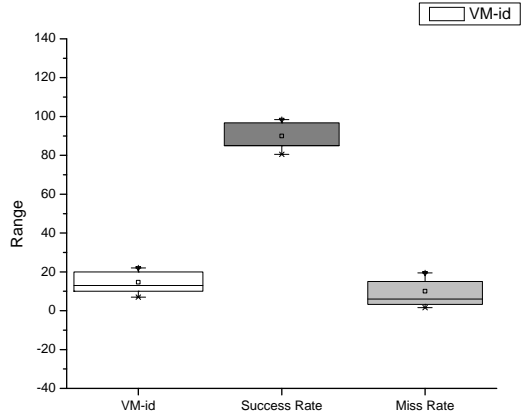}
\caption{SR and MR with respect to vm-id in cloud}
\label{figure10}
\end{figure}

$Acc_{i}$ = \begin{math}
  \begin{pmatrix}
    x_{1},&x_{2},&\cdots&,x_{n}
  \end{pmatrix}
\end{math}. There are a finite population of vm-instances so the number of accepted requests is finite as well. Hence we detect mean acceptance rate for any $X_{i}$ as,

\begin{equation}\overline{Acc_{i}} = \frac{\sum_{i=1}^{Count} Acc_{i}}{Count}\end{equation}

The value can be obtained for a large number of observations n. The variance of the delays in different ranges of Global Delays ($TBD$) and Inter-Message Time ($IMT$) is given by,

\begin{equation}{S^2(n)} = \frac{\sum_{i=1}^{Count} (Acc_{i}-\overline{Rej(Count)})^2}{Count-1}\end{equation}

Individual vm-instances were allocated for each of the SaaS applications specified and provided to cloud customers. Each customer has multiple users accessing the applications with their specific user names and passwords. For each customer, specific provenance information were collected that includes id-used to access, time and date of access, pages visited and changes made to files in terms of write operations. The provenance were detected from the SaaS level using $ObProv$ algorithm and and stored in a separate cloud server running the $Auditor$ algorithm for a period of 3 months.

The captured provenance using $ObProv$ algorithm were stored in Cloud Provenance server that as $Auditor$ algorithm implemented on the machine using Linux platform and PostgreSQL database. The $Auditor$ algorithm provides authorized access to the provenance information based on timestamp authentication, thereby enabling the prevention of undesired access that would otherwise obtain the provenance files and tamper those \cite{imran2014active}. Hence $Auditor$ algorithm ensures tamper-proofness of provenance information. Table II highlights the results of provenance access acceptance and rejections by $Auditor$ algorithm.

\section{Conclusion}

The paper aimed to ensure provenance cognition at the application layer of SaaS in the cloud. At the same time algorithm to ensure tamper-proofness of captured provenance using timestamped access control have been identified. The proposed algorithms were implemented on six real life SaaS applications running on OpenStack cloud platform. Performance analysis in terms of success and miss rates of the proposed algorithms show desirable performance of the proposed algorithms.

Empirical investigation of the proposed algorithms show an average success rate of 89.33\% that is desirable compared to industry benchmark values. The results show effectiveness of the proposed algorithms in detection of pre-specified provenance requirements. The Miss Rate (MR) of 8.66\% that is below 10\% is another desirable attribute of the proposed algorithms.

As stated earlier, the proposed algorithm detects and stores provenance in a tamper-proof format, thereby ensuring the accountability of provenance information to be used in digital forensic investigation. Association of tamper-proof provenance with web forensic data to yield more effective security intelligence is a topic of future research interest.

\section*{Acknowledgment}
This research has been supported by the University Grants Commission, Bangladesh under the Dhaka University Teachers Research Grant No-Regi/Admn-3/2012-2013/13190 and the Ministry of Science and Technology (MoST) Research Fellowship No-39.012.002.01.03.019.2013-229.

\end{document}